\documentclass[11pt,dvips]{article}

\usepackage{epsfig,times} 
\usepackage{picinpar}
\usepackage{wrapfig}
\usepackage{floatflt}
\makeatletter
\renewcommand{\section}{\@startsection{section}{1}{0in}
	{0.4\baselineskip}{0.1\baselineskip}{\Large\bf}}
\renewcommand{\subsection}{\@startsection{subsection}{2}{0in}
	{0.25\baselineskip}{-\baselineskip}{\large\bf}}
\renewcommand{\subsubsection}{\@startsection{subsubsection}{3}{0in}
	{0.1\baselineskip}{-\baselineskip}{\normalsize\bf}}
\makeatother
\begin{document}

%
%  Session and Paper Code:
\thispagestyle{myheadings}
%
%  ***INSTRUCTIONS:***  Replace `OG 9.9.9' in the command argument below
%			with your assigned session and paper code:

%
%  Session and Paper Code:
\makeatletter\newcommand{\ps@icrc}{
\renewcommand{\@oddhead}{\slshape{SH.3.2.38}\hfil}}
\makeatother\thispagestyle{icrc}
%\markright{SH.3.2.38}
%

%  Title:
\begin{center}
%
%  ***INSTRUCTIONS:***  Replace `Instructions for Preparation of Manuscript'
%			with your paper's title:
{\LARGE \bf Moon and  Sun Shadowing Observed by the MACRO Detector}

\end{center}

%  Author List:
\begin{center}
%
%  ***INSTRUCTIONS:***  Replace authors and addresses below with your own:
%
{\bf N. Giglietto $^{1}$ for the MACRO collaboration}\\
{\it $^{1}$Dipartimento di Fisica,  Universit\'a di Bari and INFN, Italy }
\end{center}

%  Abstract:
\begin{center}
{\large \bf Abstract\\}
\end{center}
\vspace{-0.5ex}
%
%  ***INSTRUCTIONS:***  Replace text below with your own abstract:
%
Using over 40 million muons collected since 1989 by the MACRO detector we have
searched for a depletion of muons coming from the direction of the Moon due to
primary cosmic rays striking the Moon.  We observe this Moon shadow in the
expected position with a statistical significance of more than 5 standard
deviations. We have analyzed the same data for an analogous Sun shadow, and
have found a signal with a significance of about 4 standard deviations.  The
Sun shadow is displaced from the Sun's position by about 0.6 degrees North in
ecliptic coordinates.  This displacement is compatible with a deflection of
primary cosmic rays due to the Interplanetary Magnetic Field in the 10-20 TeV
primary energy range which is relevant to the underground muons observed by MACRO.

%

%  Leave this line skip in place:
\vspace{1ex}

%
%  Manuscript text:
%
%  ***INSTRUCTIONS:***  Delete the next few pages of text and enter your own.  There will
%			be a warning, `STOP DELETING TEXT!!', just before the References
%			section so that the standardized Reference heading will not be
%			accidently erased.  Within the text below is an example is given
%			of a figure placement (using `picinpar').
\section{Introduction:}
\label{intro.sec}

MACRO was primarily designed to
search for magnetic monopoles and rare particles in the cosmic rays, including
high energy atmospheric neutrinos and muons from cosmic point sources 
(Ahlen et  al., 1993).
An important goal of the MACRO detector therefore is
 the recognition of point sources of
high energy cosmic rays (E$>10$~TeV) by looking for an excess of underground muons
from a particular direction of the sky and above a nearly isotropic background
of cosmic rays.
As previously suggested (Clark, 1957) the shadow in the high energy cosmic ray
flux from the directions of the Moon and the Sun could be observed and employed
to measure the angular resolution and pointing accuracy of the detector.
Clear evidence of this effect has been reported by many EAS
experiments (Aglietta et al., 1991, Alexandreas et al., 1991, Amenomori
et al., 1993, Borione et al., 1994).

 MACRO presented the first deep underground 
evidence of the Moon shadow effect (Ambrosio et al. 1999) and now
 presents evidence of the Sun  shadow due to its excellent angular 
resolution ($\approx
1^{\circ}$) and  its large collecting area and high statistics when compared to other deep
underground detectors.
Another very interesting point is that almost all cosmic ray particles are
charged and deflected by Geomagnetic (GMF) and Interplanetary (IMF) magnetic
fields. The shadows of the Moon and the Sun show these effects
 by means of displaced obscurations, whose analyses might give new
information
about these fields as well as about the cosmic-ray energy and charge
distribution. \\
\indent For this analysis we use the same technique used by Alexandreas et
al., 1991, Amenomori et al., 1993 and Borione et al., 1994.
In this work we extend the sample used for the previous Moon
 shadowing analysis (Ambrosio et al. 1999) and present
the results for the Sun shadowing effect.

\section {Event Selection:}
\label{event.sec}
The muon sample used in this study includes all events recorded between
February 1989 and  December 1998. About $45 \times 10^6$ events have been
collected over $2.6 \times 10^3$ live~days.  
The criteria used to select events for this analysis are the same 
as in Ambrosio  1999,  defined to
optimize the quality of reconstructed tracks.  The selected events are
consequently those which most accurately point back to their origin on
the celestial sphere.
These selection criteria reduce the sample size to 
$38.2 \times 10^6$ muons.

The topocentric positions of the Moon and the Sun are computed at the arrival 
time of each event in the sample, using the database of ephemerides available
from the Jet Propulsion Laboratory, JPLEPH (Standish E.M. et al.,
1995)  and include a
correction for the parallax due to MACRO's instantaneous position 
on the earth.
The muon events in two $10^\circ$ wide windows centered on the Moon and
the Sun are selected for further analysis.\\
There are $3.17 \times 10^5$ events in the Moon window that pass all cuts.
For the Sun shadow analysis there are $3.08 \times
10^5$ events collected that pass all cuts.
Twenty five background samples are generated for each run used in the
analysis.  These backgrounds are constructed by coupling the direction
of each muon in the run with the times of 25 randomly selected muons
from the same run.  The 25 background samples are then processed using 
the same procedure as the muon data sample.

\section{Shadow of the Moon:}
\label{Moonsh.sec}

\begin{figwindow}[0,r,%
{\mbox{\epsfig{file=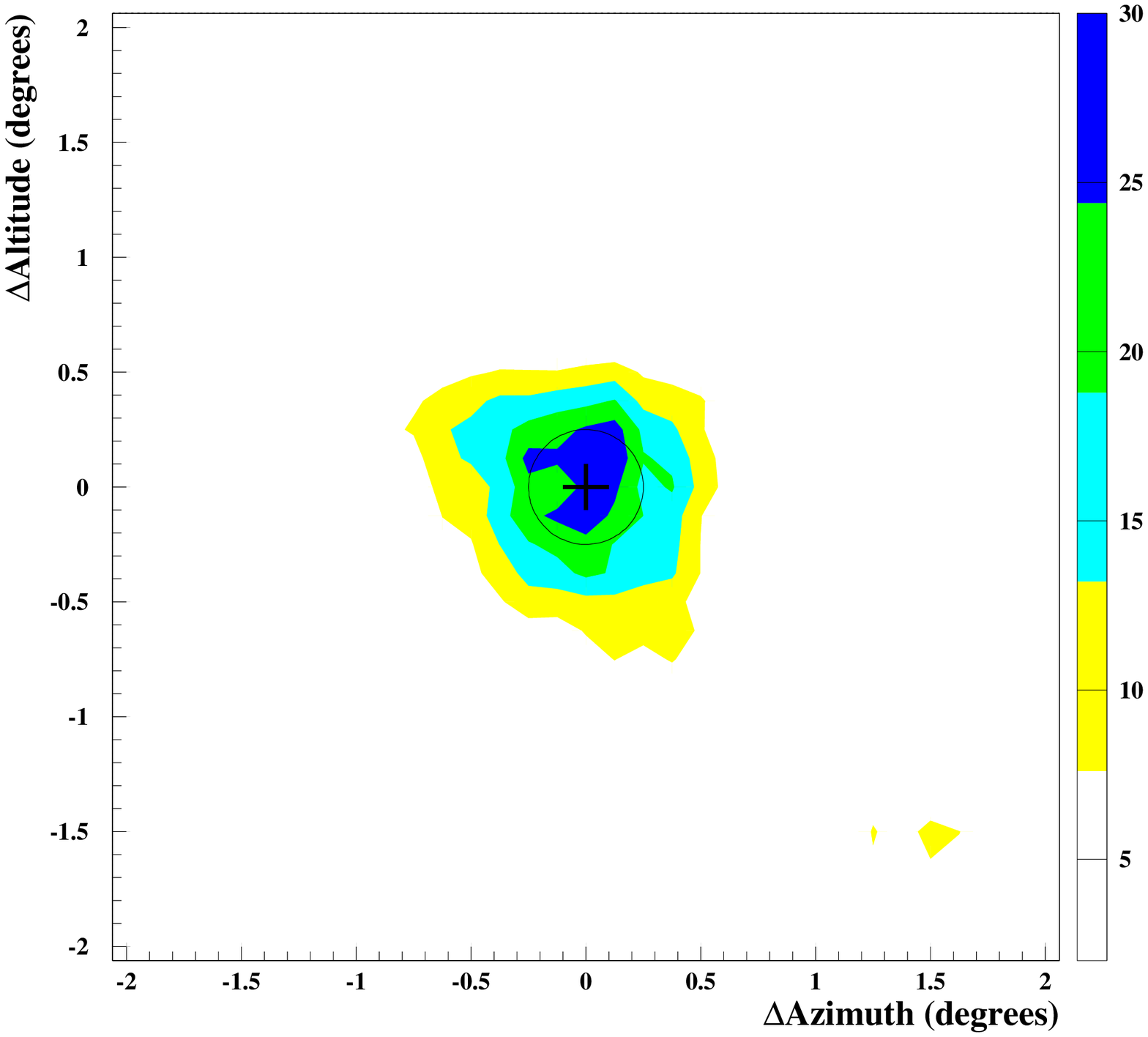,width=9.cm,height=8.0cm}}},%
{\label{Moonsh.fig} The two-dimensional distribution of $\lambda$ in bins of equal solid
angle in the Moon window. The axes are offsets from the Moon center.
 The fiducial position of the Moon, 
  at position (0,0), is marked by a $+$; a circle corresponding to the average
 lunar radius,
  $0.26^\circ$, is centered at this position.  The $\lambda$ grey scale
  is given at the right margin of the figure.  The maximum of this
  distribution  has  $\lambda = 30.0$.\\}] 
The search for the Moon shadow in a
direction-independent way and the estimate of its significance is possible 
with the maximum likelihood method of
COS-B, a technique first described in detail by
Cash W., 1979.
This method is based on {\it a priori} knowledge of the point spread function 
of the MACRO detector (PSF).
We have determined the PSF using the observed space angle
distribution of double muons divided by $\sqrt{2}$. The PSF, 
as determined from more than 10$^{6}$
muon pairs, looks strongly peaked and has a  non-Gaussian behavior
(Ambrosio et al.  1999).
To find the most likely position of the Moon, we 
compare the two-dimensional distribution of muons in the 
window centered on the Moon with the expected background events in 
the same window.  In this analysis, 
each muon event is first sorted into a grid of equal solid angle
bins ($\Delta \Omega = 0.125^\circ \times 0.125^\circ = 1.6 \times
10^{-2}$ deg$^{2}$).  
The shadowing source of strength $S_M$ 
at a fixed position $(x_s, y_s)$ 
that best
fits the data is then found by 
minimizing 

\begin{displaymath}
\lefteqn{
\chi^2 (x_s,y_s,S_M)=2 \sum_{i=1}^{n_{bin}}[ N_i^{ex}- N_i
  +N_i\ln{\frac{N_i}{N_i^{ex}}}],
}
\end{displaymath}

\noindent where the sum is over all bins in the window.  
Here $N_i$ is the number of events observed in each 
bin $i$, $N_i^{ex}$ is the expected number of events in bin $i$, and
$n_{bin}$ is the number of bins in the grid.  
This expression assumes that a Poissonian process
is responsible for the events seen in each bin.
 The expected number of
events in bin $i$ is given by 
%\begin{displaymath}
%\begin{equation}
$N_i^{ex}= N_i^{bkd} - S_M\cdot \mathcal{P}(x_s-x_i,y_s-y_i)$,
%\end{equation}
 where $N_i^{bkd}$ is the average number of background 
events at the position 
$(x_i, y_i)$, and $S_M \cdot
\mathcal{P}(x_s-x_i,y_s-y_i)$ is the number of events removed from bin
$i$ by the shadow of the Moon.  Here $\mathcal{P}(x_s-x_i,y_s-y_i)$ is the
MPSF, modified for the finite size of the Moon's disk, 
computed at the point $(x_i,y_i)$ when the shadowing source is at
$(x_s,y_s)$.  
Finally, the shadow strength
$S_M$ that minimizes $\chi^2$ is computed  
for every grid point in the window.  
This 
minimum $\chi^2 (x_s,y_s,S_M)$ is then compared with 
$\chi^2(0)$ for the 
{\it null hypothesis} that no shadowing
source is present in the window ($S_M = 0$), 
%\begin{displaymath}
%\begin{equation}
  $\lambda = \chi^2 (0) - \chi^2 (x_s,y_s,S_M)$.
%\end{equation}
%\end{displaymath}

\indent In Figure~\ref{Moonsh.fig} we show the results of this analysis 
in a window $4.375^\circ \times
4.375^\circ$ centered on the Moon.  This window has been 
divided into $35 \times 35$ cells, each having dimensions 
$0.125^\circ \times 0.125^\circ$. In this window the number of muon events is
of 14388.
In this figure, $\lambda$ is displayed in grey scale format
for every bin in the Moon window. \\
\indent Also shown  is the fiducial position of the Moon  
and a circle centered at this position corresponding to the average
lunar radius, $0.26^\circ$.
The maximum  $\lambda = 30.0$, corresponding to $5.5 \sigma$,   
is found at  the expected Moon position and this  
provides further confirmation of
MACRO's absolute pointing.  
The value of                                               
the shadow strength obtained by the likelihood method 
at this position, $S_M = 243^{+31}_{-41}$~events, agrees well
with the expected value of 215~events.  
 We have verified the properties of the $\lambda$ distribution by
constructing 71 other windows similar to the Moon window, each 
displaced from the next by $5^\circ$ in right ascension.  
For each off-source window, we have followed the 
procedure used for the Moon window in computing the expected
background.   To avoid edge 
effects associated with a source near the
edge of a window, we only have evaluated $\lambda$ for the 
central 12$\times$12 bins.
\end{figwindow}
 
\section{Angular Resolution of the MACRO Apparatus:}

If we assume that the space angle distribution of double muons is a good
approximation of PSF, the angular resolution of the apparatus can
firstly be 
estimated by this distribution as the angle $\theta_{68\%}$ of the cone that
contains the 68$\%$ of the events from a point-like source.
Using this definition of the angular resolution, %figure \ref{ang_res} gives
we obtain $\theta_{68\%}=0.8^{\circ}$.
Since  a simple Gaussian function  cannot 
fit the PSF, we cannot compute the angular resolution as the
$\sigma$ of the Gaussian function that maximizes the likelihood function for the
detection of the Moon shadow. 
Therefore we define a scale parameter $\mathcal{F}$ that rigidly scales
the PSF by the factor $\mathcal{F}$,
%\begin{displaymath}
  $\tilde{\mathcal{P}}(x_s-x_i,y_s-y_i; \mathcal{F})$
%\end{displaymath}
and  then repeat the likelihood analysis in the Moon
window for different values of $\mathcal{F}$.  We assume that the value of
$\mathcal{F}$ that maximizes $\lambda$ gives the best $\mathcal{F}$ for
computing the angular resolution.
Using these procedure  
we find that the
angular resolution is 0.86$^{\circ}$$^{-0.18^{\circ}}_{+0.26^{\circ}}$
fully consistent with the other estimation.

\section{The Sun Shadow:}
\label{sunsh.sec}
\begin{figwindow}[0,r,%
{\mbox{\epsfig{file=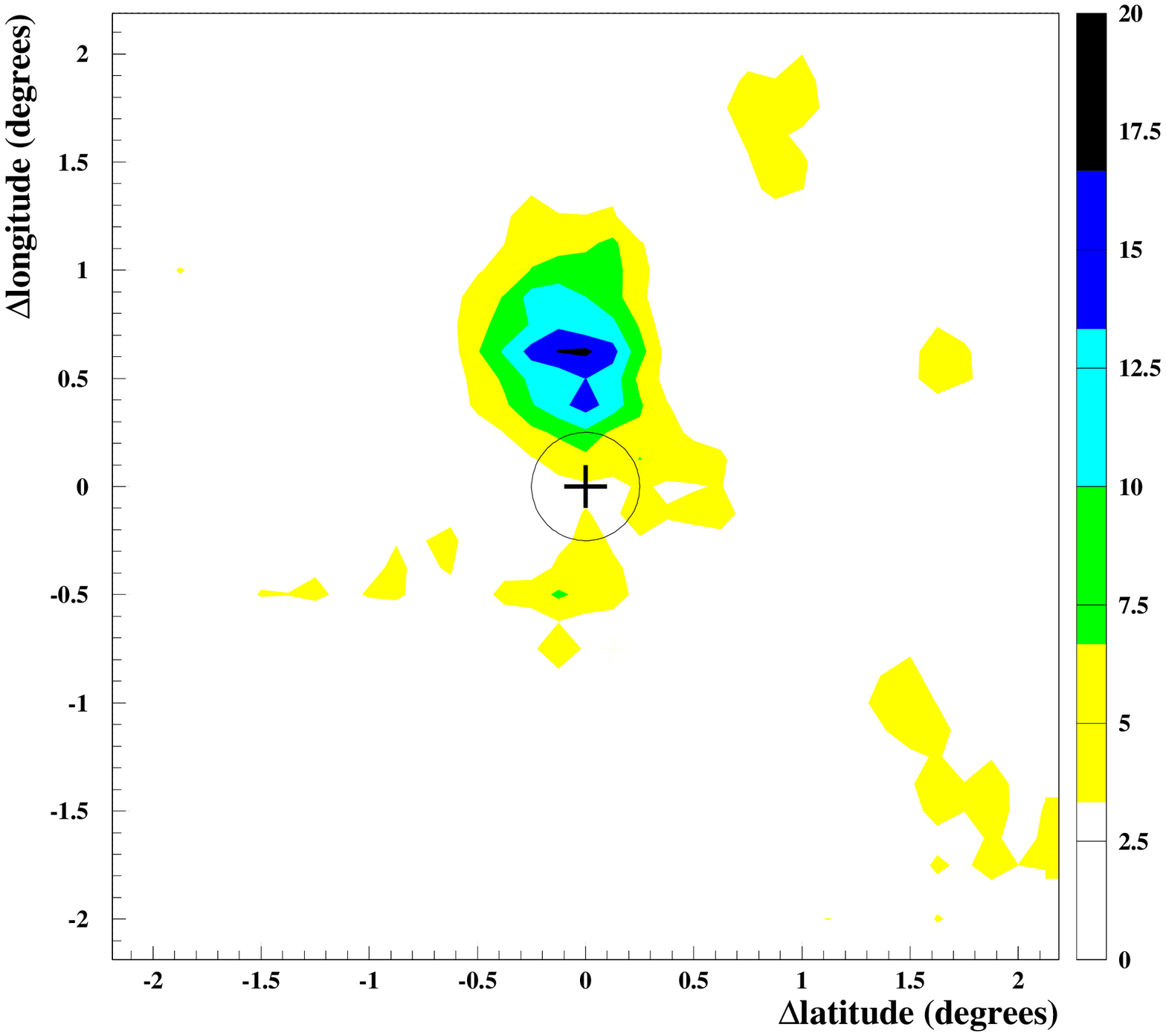,width=10.0cm,height=9.5cm}}},%
{\label{sunsh.fig} The two-dimensional distribution of $\lambda$ in bins of equal solid
angle in the Sun window. 
The axes are offsets from the Sun center using ecliptic coordinates.
 The fiducial position of the Sun, 
  at position (0,0), is marked by a $+$; a circle corresponding to the average
 solar radius,
  $0.26^\circ$, is centered at this position.  The $\lambda$ grey scale
  is given at the right margin of the figure.  The maximum of this
  distribution, $\lambda = 17.4$, is about 0.6$^{\circ}$ N
.}]
We have repeated the same analysis by using events in the Sun window. In
ecliptic coordinates (ecliptic latitude and ecliptic longitude), ~the
~maximum-likelihood\\  method gives a map of $\lambda$, in the Sun window, as shown
in Figure \ref{sunsh.fig}.
The most probable position of the center of the deficit is found at
(0$^\circ$,0.625$^\circ$) with a $\lambda$=17.4 corresponding to a significance
of 4.2 $\sigma$. The probability to find random fluctations having a
$p(\chi^2 \ge 17.4)$ is less than 10$^{-5}$, hence the observed 
Sun's shadow is significant.

\indent The large displacement of the shadow from the apparent position of the Sun
is likely due to the combined effect of the magnetic field of the Sun and the
geomagnetic field. However a quantitative interpretation of the Sun's shadow in
connection with
the effect of the solar magnetic field is difficult, because during almost ten years
of MACRO data acquisition, the effect of the IMF is averaged over its changes
due to the variable inclination and over the yearly variation of solar activity
with a period of 22 years. Moreover the IMF has a sector structure with the
magnetic field direction reversed so that in some sectors the magnetic field
points inwards and in others outwards and the neutral sheet separating the two
regions is inclined to the equatorial plane.
As the earth is within 7$^\circ$ of the solar equatorial plan, the polarity of
the IMF will change several times during the Sun rotation of 27 days on its
axis.\\
\end{figwindow}
A careful selection of the data as a function of IMF polarity configuration,
as well as higher statistics, could be used to obtain new 
information about the
IMF.\\
%
%The effect of the magnetic fields on the curvature of cosmic rays could be used
%in the search of antimatter in the Universe and particularly in the measuring of
%the antiproton-proton ratio at TeV energies. Antiproton and proton 
%in primary cosmic rays are deflected in opposite positions by the IMF thus they
%should give two opposite deficits with regard to the real position of the Sun: 
% $\bar{p} / p$ ratio is given by the ratio of the two deficits or the significances
%of  the two shadows (Amenomori et al., 1995).\\
%The significance of  $\bar{p}$ deficit does not exceed 1 $\sigma$ so the upper limit
%of $\bar{p} / p$ ratio is roughly given by 1/4.2=24 $\%$
% being 4.2 $\sigma$ the significance of the $p$ Sun shadow.

\section{Conclusions:}

The MACRO detector, operational since
February~1989, has collected a muon sample of about 40~million events.
Using this sample we have observed the Moon and Sun shadows in the cosmic
ray sky at primary energies $\approx$ 10 - 15 TeV. In the deficit analysis, we 
find an event deficit
around the Moon of about $5.0 \sigma$ significance.
  With a maximum likelihood analysis, we confirm the detection of
the Moon's shadow with a significance of $5.5 \sigma$.  Our estimate of the
angular resolution is
$\theta_{68\%}$ =0.86$^{\circ}$$^{-0.18^{\circ}}_{+0.26^{\circ}}$.
These results characterize MACRO as a muon telescope by confirming
MACRO's absolute pointing ability and by quantifying its angular
resolution.  This investigation shows that the MACRO detector has the
capability of detecting signals from point-like sources by detecting
secondary underground muons .
The analysis of the deficit in the Sun direction and in ecliptic coordinates gives
a Sun shadow with a significance of 4.2 $\sigma$.
The shadow of the Moon and the Sun is shifted due to the curvature
of cosmic rays by GMF and IMF at the primary energy of 10-20 TeV
relevant to MACRO.

%
%  **FIGURE EXAMPLE**  This uses the package `picinpar', which is available at
%			the ICRC web site.

%
%
%
%  STOP DELETING TEXT!!
%  STOP DELETING TEXT!!
%  STOP DELETING TEXT!!
%  STOP DELETING TEXT!!
%  STOP DELETING TEXT!!
%  STOP DELETING TEXT!!
%  STOP DELETING TEXT!!
%  STOP DELETING TEXT!!
%  STOP DELETING TEXT!!
%
%
%
%  References: (DO NOT ALTER NEXT 4 LINES)
\vspace{1ex}
\begin{center}
{\Large\bf References}
\end{center}
%
%  ***INSTRUCTIONS:***  Enter your references alphabetically following the format
%			of the example citations below.
Aglietta M., et al., XXII ICRC Dublin 1991, 1, 404  and also 
Astroparticle Physics 1995, 3,1.\\
Alexandreas et al.,  1991, Phys.Rev. D 43, 1735 \\
%\bibitem{tibet1}Amenomori et al., {Shadowing of Cosmic Rays by the Sun Near
%Maximum or at the Declining Phase of solar activity}, Astrophysical~Journal,
%{\bf 464}, 954 (1996)
Ambrosio M. et al, 1999 Phys.Rev.D, 59, 12003-1.\\
Amenomori et al., 1993 Phys.~Rev.~D 47,2675 \\
Amenomori et al., 1995 Proc. ICRC (Roma), 84.\\
Borione et al., 1994, Phys.~Rev.D 49, 1171 \\
Cash W.,  1979, ApJ 228, 939.\\
Clark, G.W., 1957 Phys. Rev. 108,450 \\

\end{document}